\documentstyle[buckow]{article}
\newcommand{\beq}[1]{\begin{equation}\label{#1}}
\newcommand{\eeq}{\end{equation}}
\newcommand{\beqar}[1]{\begin{eqnarray}\label{#1}}
\newcommand{\eeqar}{\end{eqnarray}}
\begin {document}
\begin{flushright}
NTZ  4/99 \\
YERPHY 1529(2)-99 
\end{flushright}
\def\titleline{
 High-energy scattering in gauge theories
\newtitleline
and integrable spin chains \footnote{\sl Contribution presented
 by R.K. to the 32nd Ahrenshoop Symposium on the 
Theory of Elementary Particles, Buckow, September 1998}}

\def\authors{
D.R. Karakhanyan\1ad ,  R.~Kirschner\2ad
}
\def\addresses{
\1ad
Yerevan Physics Institute\\
Alikhanyan Br. street \\
Yerevan, Armenia \\

\2ad
Naturwissenschaftlich-Theoretisches Zentrum  \\
und Institut f\"ur Theoretische Physik, Universit\"at Leipzig
\\
Augustusplatz 10, D-04109 Leipzig, Germany }

\def\abstracttext{
In the leading log approximation and at large $N_C$ 
the interaction of two fermionic and one gluonic reggeons
is described by an integrable system corresponding to an 
open spin chain.  
}
\large
\makefront
\section{Reggeons in gauge theories}
\noindent
In the Regge limit, where the energy is large compared to the masses and
momentum transfers, the perturbative expansion of the scattering
amplitudes is most conveniently written in terms of the exchange of
reggeons in the $t$-channel. The reggeons interact with the scattering
near mass shell quanta and with each other via effective vertices. It is
appropriate to Mellin transform the energy dependence of the
amplitudes and to consider the resulting partial waves in the complex
angular momentum ($j$) plane. The reggeons correspond to poles in $j$
moving with the tranferred momentum \cite{Lrev}.
The scattering amplitude of gauge-group singlet particles is represented
as a convolution of two impact factors and the reggeon Green function.
The impact factors describe the coupling of the in-and outgoing (in
forward and backward directions, respectively) particles to the exchanged
reggeons. Our discussion is restricted to the reggeon Green function
describing the exchange of the interacting reggeons.

The high-energy effective action \cite{eff} describes the interaction of the
scattering partons and the reggeons. This action is derived from the
underlying action by integrating over modes of the fields, that
appear neither in the scattering nor in the exchanged quanta. On the
tree level the reggeons are the modes of the original fields
corresponding to the exchange in the Regge kinematics. Due to loop
corrections the corresponding pole in the $j$-plane moves with the
momentum transferred by the reggeon. Actually, the trajectory functions
and the bare reggeon interactions have to be regularized to avoid
infrared divergences, which cancel however in the scattering amplitudes
of gauge group singlet objects.

We disregard the case of changing number of reggeons during the
exchange. The dependence on the longitudinal (time and direction of the
collision axis) dimensions is trivial. At large $N_C$ also the gaue
group part of the inteaction becomes trivial. The interaction is
pairwise. The interaction of two reggeons at impact parameters
(tranverse positions) $x_1$ and $x_2$ (paramentrized as complex numbers)
is described by hamitonian operators, which can be written as a sum of
two terms, the first involving only the holomorphic coordinates and
derivatives and the second involving only the anti-holomorphic ones.
The lowest energy eigenvalues are directly related to the right-most
Regge singularities generated by the (two- or multi-) reggeon exchange.
The reggeons are characterized by conformal weights ($\Delta, \overline
\Delta) $ which are (0,0) for gluonic reggeons and (${1  \over 2 }, 0$ )
or ($0,  {1 \over 2 }$) for fermions, depending on helicity \cite{Lop,RK94}.
We restrict the discussion to the holomorphic part keeping in mind the
condition of single-valuedness of the reggeon Green function in the
transverse plane. We denote the holomorphic part of the hamiltonian by
$H_{12}^{\Delta_1, \Delta_2} $, with $\Delta_1, \Delta_2$ being the
holomorphic weights of the two reggeons and the lower indices being the
short-hand notation for the dependence on the holomorphic coordinates
$x_1, x_2$ and  derivatives $\partial_1, \partial_2$, $ (x_{12} = x_1
- x_2) $.
\beqar1
H_{12}^{\Delta_1,\Delta_2} \ = \
\partial_1^{-1 + 2 \Delta_1} \ \ln x_{12} \ \partial_1^{1 - 2\Delta_1 }  +
\partial_2^{-1 + 2 \Delta_2} \ \ln x_{12} \ \partial_2^{1 - 2 \Delta_2 } +
 \ln \partial_1 \partial_2
- 2 \psi (1)   = \cr
2 \ln x_{12}  + 
x_{12}^{1 - 2 \Delta_1}  \ \ln \partial_1  \  \ x_{12}^{-1 + 2 \Delta_1} +
x_{12}^{1 - 2 \Delta_2}  \ \ln \partial_2  \  \ x_{12}^{-1 + 2 \Delta_2}
- 2 \psi (1).
\eeqar
The conformal (holomorphic {\sl M\"obius} ) transformations of the wave
function of the reggeon $\Delta_1$ at $x_1$ are generated by
\beq2
M_1^{-} \  \ = \  \partial_1,\ \  M_1^{0} \ = \ x_1 \partial_1 + \Delta_1,
\ \  M_1^{+} \ = \ x_1^2 \partial_1 + 2 x_1 \Delta_1 .
\eeq
It is not difficult to check that  the hamiltonian is
symmetric, i.e. commutes with $M_{12}^{a} =M_1^{a} + M_2^{a},\ 
a= \pm, 0$. Therefore there is a representation of this operator as a
function of the Casimir operator
\beq3
C_{12}^{\Delta_1, \Delta_2} \ = \ -  M_{12}^{0 \ 2} +
{1 \over 2} ( M_{12}^{+} M_{12}^{-} + M_{12}^{-} M_{12}^{+} ) .
\eeq
We have indeed
\beq4
H_{12}^{\Delta_1, \Delta_2} \ = \
 - {1 \over 2}  ( \chi_{(\Delta_1 - \Delta_2)} ( C_{12}^{\Delta_1, \Delta_2}
)
+ \chi_{(\Delta_2 - \Delta_1)} ( C_{12}^{\Delta_1, \Delta_2} ) ) .
\eeq
Writing the argument as $C = m(1-m)$ the functions $\chi_{\delta} (C)$
can be expressed in terms of the logarithmic derivative of the $\Gamma $
function
\beq5
\chi_{\delta} (C) \ = \ 2 \psi (1) - \psi (m + \delta ) - \psi (1 - m +
\delta ).
\eeq
The conformal symmetry can be established also by looking more closely
at the two forms given in (1) and observing that the formal
transposition of these operators $H^T$ can be obtained from  $H$ by
similarity transformations. For example, in the case of $\Delta_1 =
\frac{1}{2}, \Delta_2 = 0 $ we have
\beqar6
(H^{(\frac{1}{2}, 0)} )^T \ = \ \partial_2 \ H^{(\frac{1}{2}, 0)} \
\partial_2 ^{-1} \ =
{\cal P}_{12} x_{12}^{-1} \
H^{(\frac{1}{2}, 0)}  \ x_{12} {\cal P}_{12} .
\eeqar
${\cal P}_{12} $ permutes the points 1 and 2. By comparison we obtain
that there is an operator commuting with $ H^{(\frac{1}{2}, 0)} $:
\beq7
[A_{12}, H_{12}^{(\frac{1}{2}, 0)} ] \ = \ 0, \ \
A_{12} = {\cal P}_{12} x_{12} \partial_2.
\eeq
This result is directly related to conformal symmetry, because $A_{12}$
is essentially the square root of the Casimir operator:
$A_{12}^2 = C_{12}^{(\frac{1}{2}, 0)} - \frac{1}{4} $.

\section{Three-reggeon exchange}
\noindent
The symmetry which has been exhibited in the representation (4) allows
to obtain easily the eigenvalues and eigenfunctions of the two-reggeon
system. The two-reggeon Green function can be decomposed into these
eigenfunctions. For the eigenvalues we have just to substitute in (4) 
$C = m(1-m), m = \frac{1}{2} + n + i \nu$ , $n$  integer and $\nu$ real.
The contribution of multi-reggeon exchange is physically relevant not
only as unitarity correction to the two-reggeon exchange. The quantum
numbers in some channels cannot be transferred by just two gluons or by
quark - antiquark. Much work has been done for the case of multiple
exchange of gluonic reggeons. The problem has been related to integrable
spin chains \cite{Lpis,FK}.
Due to the large $N_C$ approximation one is led to a closed
chain with nearest neighbour interactions.
In the case of three gluonic reggeons (odderon channel) after serious
efforts the leading eigenvalue has been calculated \cite{JW,L98}.

We are studying the reggeon exchange with fermions included. As the
first non-trivial case we have considered the quark - antiquark - gluon
exchange. If the fermions have opposite helicities the holomorphic part
corresponds to a one-dimensional three-body system with the conformal
weights
$\Delta_1  = \frac{1}{2}, \ \Delta_2 = \Delta_3 = 0 $.
We show that the system is integrable by constructing an additional
operator, besides of the conformal generators $M_1^{a} + M_2^{a} +
M_3^{a} $, commuting with the latter and with the hamiltonian
\beq8
H_{123} \ = \ H_{12}^{(\frac{1}{2}, 0)}  + H_{23}^{(0,0)}.
\eeq
For the fermions in the fundamental representation the large $N_C$
approximation results in an open chain with nearest neighbour
interaction and with the fermions at the end points. Including adjoint
fermions we would encounter also closed inhomogeneous chains, which is a
case essential different from the one considered.

We know two ways of constructing the additional integral of motion. The
first one is a simple generalization of the transposition argument
presented above.
Indeed we observe
\beqar9
H_{123}^T \ = \ \partial_2 \partial_3 \ H_{123} \ (\partial_2
\partial_3)^{-1} =
{\cal P}_{123} x_{23}^{-1} x_{12}^{-1} \ H_{123} \ x_{23} x_{12}
{\cal P}_{123}.
\eeqar
${\cal P}_{123} $ permutes the points $ 1, 2, 3$ into $3, 2, 1$.
Therefore we have the commuting operator $A_{123}$ :
\beq1
[A_{123}, H_{123}] \ = \ 0,  \ \
A_{123} = {\cal P}_{123}  \ x_{12} x_{23} \partial_2 \partial_3.
\eeq

\section{Open spin chain}
\noindent
In the case of only gluons the multi-reggeon system is described by a
closed quantum spin chain with the non-compact $sl(2)$ representations
of weight $\Delta = 0$ at each site. We associate a {\sl Lax } matrix to each
site:
$L_i (\theta) = \theta  I^{(2)} + \sigma^{0} M_i^{0} + \sigma^{+} M_i^{-}
- \sigma^{-} M_i^{+} $.
It obeys the usual {\sl Yang - Baxter } - relation with the $4 \times
4$ $R$-matrix $ R_{12} (\theta) = \theta I^{(4)} + \xi {\cal P}_{12} $.
 This relation generalizes to
products $ \prod_{i=1}^{N} L_i (\theta + \delta_i ) = T (\theta) $. It
follows that the trace $  {\rm tr} \ T(\theta) $ generates a complete set of
commuting operators, which are just the ones needed for the integrable
closed homogeneous chain \cite{Lpis,FK}.

In the case of open chains the boundaries impose additional conditions
on the integrable interaction \cite{ChS}
\beqar1
R_{12} (\theta_1 - \theta_2) \ {\cal T}(\theta_1 ) \otimes I^{(2)} \ 
R_{12} (\theta_1 + \theta_2) \ I^{(2)} \otimes {\cal T} (\theta_2) = \cr
 {\cal T}(\theta_2 ) \otimes I^{(2)}
\ R_{12} (\theta_1 + \theta_2) \ I^{(2)} \otimes {\cal T} (\theta_1)
\ R_{12} (\theta_1 - \theta_2) .
\eeqar
and another condition related to this one by crossing.
Given a matrix $K_- $ obeying (11) when inserted for ${\cal T}$ 
and another matrix $K_+$ obeying the
crossing relation one finds that
${\cal T}(\theta) = T(\theta) K_- T^{-1} (-\theta) $
obeys (11) too, provided $T(\theta) $ obeys the ordinary  {\sl Yang -
Baxter} - relation. Furthermore, it follows that the trace ${\rm tr} ({\cal
\ T}(\theta) K_+)$ generates mutually commuting operators.
Whereas the generating function ${\rm tr} \ T(\theta)$ for the closed chain can
be visualized by drawing the closed chain and associating the {\sl Lax}
matrices with each site, the generating function ${\rm tr} \ ( T (\theta) K_- 
T^{-1}(-\theta)  K_+)$ is visualized as the closed chain built from the considered open
chain together with its mirror image and with the matrices $K_{\pm}$ 
standing for the mirror.

We obtain the integrals of motion for our three-reggeon problem for the
simple choice $K_{\pm} = I^{(2)}$. We have
\beq2
T(\theta) =
L_1^{(\Delta  = \frac{1}{2}) } (\theta + \delta_1) \
L_2^{(\Delta  = 0 )} (\theta + \delta_2) \
L_3^{(\Delta  = 0 )} (\theta + \delta_3) .
\eeq
The trace 
${\rm tr} ( T(\theta) T^{-1}(-\theta) )$, e.g. with $\delta_1 = \delta_2 =
\delta_3 = \frac{1}{2}$, decomposes into the operators
$A_{123}^2$ (10), $ C_{123} = C_{12} + C_{23} + C_{31} $ and $C_{23}$ (3)
with linear independent functions of $\theta$ as coefficients.
In this way we have checked that the considered three-reggeon system is
properly described by an integrable open spin chain. Now the generalization
to
multi-reggeon systems with quark and antiquark and an arbitrary number
of gluonic reggeons is straightforward.
\vskip0.5cm
\noindent
{\large \bf Acknowledgements}

\noindent
The work has been supported in part by Deutsche Forschungsgemeinschaft,
grant No. Ki~623/1-2.

\end{document}